\documentclass[manuscript, screen]{acmart}

\def\markup{0}  
\if\markup 1
\newcommand{\rv}[1]{{\leavevmode\color{black}#1}}
\newcommand{\rw}[1]{{\leavevmode\color{black}#1}}

\newcommand{\h}[1]{{\leavevmode\color{black}#1}}
\newcommand{\camera}[1]{{\leavevmode\color{blue}#1}}
\else
\newcommand{\rv}[1]{#1}
\newcommand{\rw}[1]{#1}

\newcommand{\h}[1]{#1}
\newcommand{\camera}[1]{#1}
\fi

\AtBeginDocument{%
  \providecommand\BibTeX{{%
    \normalfont B\kern-0.5em{\scshape i\kern-0.25em b}\kern-0.8em\TeX}}}

\copyrightyear{2022} 
\acmYear{2022} 
\setcopyright{acmlicensed}\acmConference[ASSETS '22]{The 24th International ACM SIGACCESS Conference on Computers and Accessibility}{October 23--26, 2022}{Athens, Greece}
\acmBooktitle{The 24th International ACM SIGACCESS Conference on Computers and Accessibility (ASSETS '22), October 23--26, 2022, Athens, Greece}
\acmPrice{15.00}
\acmDOI{10.1145/3517428.3544820}
\acmISBN{978-1-4503-9258-7/22/10}

\usepackage{colortbl}
\usepackage{graphicx}
\usepackage{enumitem}
\usepackage{CJKutf8}
\usepackage{multirow}

\begin{document}


\title[Understanding Banking Practices and Challenges Among Older Adults in China]{``I Used To Carry A Wallet, Now I Just Need To Carry My Phone'': Understanding Current Banking Practices and Challenges Among Older Adults in China}


\author{Xiaofu Jin}
\affiliation{%
  \institution{IIP(Computational Media and Arts)}
  \institution{The Hong Kong University of Science and Technology}
  \city{Hong Kong SAR}
  \country{China}
}
\email{xjinao@connect.ust.hk}

\author{Mingming Fan}
\authornote{Corresponding Author}
\affiliation{
  \institution{Computational Media and Arts Thrust}
  \institution{The Hong Kong University of Science and Technology (Guangzhou)}
  \city{Guangzhou}
  \country{China}
}
\affiliation{
  \institution{Division of Integrative Systems and Design}
    \institution{Department of Computer Science and Engineering}
  \institution{The Hong Kong University of Science and Technology}
  \city{Hong Kong SAR}
  \country{China}
}
\email{mingmingfan@ust.hk}

\renewcommand{\shortauthors}{Jin and Fan}

\begin{abstract}

Managing finances is crucial for older adults who are retired and may rely on savings to ensure their lives’ quality. As digital banking platforms (e.g., mobile apps, electronic payment) gradually replace physical ones, it is critical to understand how they adapt to digital banking and the potential frictions they experience. We conducted semi-structured interviews with 16 older adults in China, where the aging population is the largest and digital banking grows fast. We also interviewed bank employees to gain complementary perspectives of these help givers. Our findings show that older adults used both physical and digital platforms as an ecosystem based on perceived pros and cons. Perceived usefulness, self-confidence, and social influence were key motivators for learning digital banking. They experienced app-related (e.g., insufficient error-recovery support) and user-related challenges (e.g., trust, security and privacy concerns, low perceived self-efficacy) and developed coping strategies. We discuss design considerations to improve their banking experiences.

\end{abstract}



\maketitle

\section{Introduction}

Banking, as an indispensable service of daily life, is undergoing unprecedented changes under the trend of the digital economy. With the increasing penetration of high-speed internet and electronic devices (e.g., tablets, smartphones), digital banking has become increasingly pervasive~\cite{aladwani2001online, OnlineShopping2021, von2004electronic,The2020M23:online}. 
In addition to banking websites and mobile banking apps, electronic payment services (e.g., AliPay \cite{Alipay}, ApplePay \cite{ApplePay}, Google Pay \cite{googlepay}, PayPal \cite{paypal}, Venmo \cite{venmo}, WeChat Pay \cite{Wechatpay}, Yunshanfu \cite{Yunshanfu}, ZellePay \cite{zelle}) and virtual (i.e., online-only) banks (e.g., Netbank \cite{NetBank}, Webank \cite{WeBank}, Monzo~\cite{Monzo}) recently emerged as alternative digital banking platforms. 
Moreover, this trend has been accelerated by the COVID-19 pandemic. Many banks have shut down their physical branches and replaced them with more digital banking platforms~\cite{Hongkong, SwitzerlandBanks, USBanks,BBCNews}. 

As face-to-face interactions in physical banks have been gradually replaced by digital user interfaces on websites and mobile apps, researchers have investigated people's experiences and attitudes toward online and mobile banking~\cite{Laforet2005, Hua2008, yang2009comparative} and digital payment~\cite{liebana2020mobile,zhao2014exploring}. 
However, such studies primarily focused on young adults. Compared to young adults, older adults tend to use technologies to a lesser extent \rv{and feel more reluctant to adopt new technologies~\cite{olson2011diffusion,czaja2006factors, vaportzis2017older}. They also tend to encounter more difficulties when using new technologies in general due to factors such as age-related declines (e.g., cognitive, physical)~\cite{adoption,adoption2,harris2016consumer}, generation/cohort effects~\cite{adoption}, digital divide/barriers~\cite{ordonez2011elderly, Choudrive2018}, and fewer educational opportunities to keep up with the technology~\cite{Pikna2018Information}}.
Consequently, older adults might face more challenges when adopting digital banking. 
 
 On the other hand, compared to young adults, older adults have accumulated more experiences with a variety of technologies over decades and may have different criteria for ``good technology''. 
 As a result, understanding what older adults have to say about technology would be beneficial for designing more inclusive technology, not just for themselves but also for everyone. After all, aging is a process that everyone is experiencing~\cite{Knowles2021TheHarm}. 
 Thus, it is critical to understand older adults' banking experiences to improve the accessibility of emerging digital banking platforms. 

Researchers have already begun to explore older adults' digital banking experiences recently~\cite{gatsou2017seniors, omotayo2020digital, peral2019self,Msweli2020Enablers, trust16, fear, Choudrive2018}. While informative, these studies primarily focused on \textit{specific aspects} of digital banking, such as the user experience of bank websites on desktop computers~\cite{gatsou2017seniors}, strategies to build trust in internet banking~\cite{trust16}, and the effect of self-efficacy and anxiety on internet banking~\cite{peral2019self}. 
A recent survey study took a step further to understand older adults' overall banking practices in light of recent technological development and found that although the majority of older adults still used physical banks, a small percentage of them started to adopt digital banking and digital payments~\cite{Xiaofu21}. 
However, this survey study mainly provided a \textit{quantitative} overview of older adults' banking practices, and it remains largely unknown \textit{why and how older adults choose to use different baking platforms and the challenges they encounter}. 

Inspired by this line of work, we took a step further to explore the following research questions (RQs):
\begin{itemize}
\item {RQ1}: \rw{How do older adults use physical and digital banking platforms? What are the perceived pros and cons of these banking platforms?}
\item {RQ2}: What are older adults' motivations for learning digital banking?
\item {RQ3}: What are the challenges that older adults encounter when using digital banking platforms?
\end{itemize}

To answer RQs, we conducted in-depth semi-structured interviews with 16 older adults living in five cities in China. China has the largest older adult population~\cite{world2015china,mukhtar2020psychological} and has been undergoing a fast growth in digital banking. For example, China is a world leader in the adoption of contactless mobile payments with 81.1\% usage penetration~\cite{Hua2008, GlobalMP2019} and has the highest growth rate of electronic-payment transactions among all countries~\cite{The2020M23:online}. Under this rapid transition to digital banking, older adults in China have opportunities to experience both physical and digital banking platforms and may encounter challenges associated with various digital banking platforms. 

\rv{Furthermore, we also conducted semi-structured interviews with bank employees to gain an understanding of older adults' banking experiences from their complimentary perspectives as help-givers. These bank employees had direct interactions with older adults and accumulated experiences when assisting older adults with their banking needs.} 

 Our findings show that older adults used both physical (e.g., bank counters, ATMs) and digital banking (e.g., website, mobile bank apps, electronic payment, virtual bank apps) platforms. They performed different types of transactions on different platforms based on their perceived pros and cons.
Moreover, our study uncovered three motivations for learning digital banking, which were perceived usefulness, self-confidence, and social influence(e.g., being motivated by people in social circles). Furthermore, our study found that older adults encountered app-related challenges (e.g., ambiguous affordance, low information scent, insufficient error recovery support, lack of feedback) and user-related challenges (e.g., anxiety caused by intangibility, trustworthiness, security, and privacy, low perceived self-efficacy and memory concerns) when using digital banking platforms.  

We found that although older adults encountered challenges, they showed a strong willingness to learn and apply their accumulated experiences and knowledge to tackle the problems first before seeking help from ones whom they trust. 
Taken all together, we further discuss the implications of our key findings and propose design considerations and future work for assisting older adults with overcoming challenges in using digital banking \rv{and for supporting help-givers to better assist older adults.} 
In sum, we make the following contributions based on both older adults' and bank employees' perspectives:
\begin{itemize}
\item An understanding of how older adults use physical and digital banking platforms and their perceived pros and cons of these platforms;
\item An understanding of the motivations to learn digital banking;
\item An understanding of banking challenges and design considerations to improve older adults' banking experiences.
\end{itemize}

\section{Background and Related Work}
\subsection{Banking Trend and Technology Adoption for Older Adults}
Banking services are increasingly digitized with the development of information and communication technology (ICT) \cite{Faris2019}. 
At the same time, physical banks also constantly reduce the number of branches and staff to save money \cite{Mckinsey2019}. 
While digital banking could bring convenience to people who are adept at digital technology, such a rapid technological shift may pose challenges to people who are accustomed to traditional physical banking, such as older adults.

Prior research suggested that older adults tend to adopt new technologies (e.g., computers, internet, tablets) slower and may be less likely to use technologies in general than young adults~\cite{czaja2006factors,olson2011diffusion,vaportzis2017older}. 
Olson et al. conducted a survey study with 430 younger adults and 251 older adults
and found that older adults tend to be frequent users of long-standing technologies (e.g., telephone) and less frequent users of more recent technologies (e.g., Internet, ATMs)~\cite{olson2011diffusion}. Digital banking was also found to be not as popular among older adults as it was among the young \rw{in Asia}~\cite{Mckinsey}. 

Possible reasons for this age difference could be the age-related health declines and the digital divide among older adults. Age-related perception declines like vision impairments may make it more difficult for older adults to perceive small icons when using technology devices \cite{czaja2009information, mohadisdudis2014study}, and age-related physical declines like reduction in fine motor skills can cause older adults to encounter more motor issues such as tapping and scrolling when using mobile apps \cite{yu2020maps}, and age-related cognitive issues, such as the reduced speed of learning and memory difficulties, may also slow down their learning of digital technologies \cite{Delello2017}.  \rw{Moreover, older adults were exposed to a different set of technological products when they were in the workforce compared to younger generations~\cite{ordonez2011elderly, Choudrive2018}, which could potentially contribute to the digital divide between older and young adults}. 

As a result, as an emerging new technology, digital banking might impose new challenges for older adults, \rw{especially under the current COVID-19 pandemic, people in many countries are advised to avoid using in-person banking}. To ensure the inclusiveness of digital banking for older adults, it is critical to understand how older adults perform banking activities on digital banking platforms and the associated challenges. 

In this work, we seek to explore this question with older adults in the context of China.
China has the world's largest and fastest-growing aging population \cite{world2015china,mukhtar2020psychological}. It is estimated to have 280 million older adults by 2025, which will represent about one-fifth of its total population \cite{world2015china,mao2020china}. At the same time, the development of the digital banking and e-commerce market in China is among the fastest-growing in the world with a volume of 1.94 trillion USD in 2019 \cite{eMarketer2019, Hua2008}, and has almost become a cashless economy with the fastest growing electronic payment \cite{lipton2016digital, KPMG2020China,The2020M23:online}. As a result, older adults in China have been experiencing the fastest shift from traditional physical banking to digital banking and may experience more challenges when learning and using new banking platforms.     
By studying the banking practices and challenges that older adults in China experience, we hope to reveal design opportunities to better improve the banking experience for the aging population.

\subsection{Older Adults' Banking Practices}
Physical banks (e.g., bank counters) are commonly used by older adults \cite{abood2015can,AgeUK2016, harris2016consumer,omotayo2020digital}, and older adults were found to prefer to visit bank branches over conducting banking transactions online \cite{mattila2003internet, Camilleri2017, omotayo2020digital}.
In-person customer service offered in physical banks was rated as one of the top desired services from financial institutions among older adults~\cite{abood2015can}. Despite the familiarity with physical banks, older adults complained about the inconvenience associated with visiting bank branches, such as the long wait time~\cite{omotayo2020digital, Xiaofu21}. 

The automated teller machine (ATM) was a technological innovation to traditional banking products~\cite{Choudrive2018,BBC2007}, and the adoption rate of ATMs among older adults has been slowly increasing over the past decades~\cite{Gilly1985,zeithaml1987characteristics, Roger1996,Darch2004,o2008understanding}. Nonetheless, many older adults still do not use ATMs~\cite{Xiaofu21}. One critical reason was that older adults felt uncomfortable and less in control of their finances when using an ATM~\cite{Darch2004}. O’Brien et al. further identified more factors influencing ATM adoption among older adults, which include usefulness, compatibility, complexity, technology generation, and relative advantage of a technology~\cite{o2008understanding}. \rv{Compared to traditional ATMs, newer versions of self-service banking machines, such as Cash Recycling System (CRS) machines, could handle more banking transactions. These CRS machines typically have a big touchable screen and can handle many banking transactions. In this paper, we treat all of them as ATMs.}

Digital banking platforms, such as bank websites and mobile banking apps, have recently emerged as alternatives to physical banks and ATMs \cite{Faris2019}. Researchers investigated whether and to what extent older adults adopted digital banking platforms (e.g., ~\cite{gatsou2017seniors, omotayo2020digital, peral2019self,Msweli2020Enablers, trust16, fear, Choudrive2018}).
\rw{Compared to young adults, older adults tend to use digital banking to a lesser extent and preferred to use the traditional banking platforms (e.g., telephones, physical banks)~\cite{Mckinsey, olson2011diffusion, omotayo2020digital}}. 
\rv{Furthermore, researchers investigated the factors to affect older adults' adoption of digital banking platforms, including the user experience of banking websites~\cite{gatsou2017seniors},  trust~\cite{trust16}, fear~\cite{fear}, and self-efficacy and anxiety \cite{peral2019self}.}

In sum, prior research primarily focused on the adoption of \rw{\textit{a particular type}} of banking platforms (e.g., physical banks \cite{abood2015can}, ATMs \cite{Roger1996,zeithaml1987characteristics,Gilly1985,Darch2004,o2008understanding}, or digital banking platforms \cite{mattila2003internet,Msweli2020Enablers,trust16,Choudrie2018,chawla2018moderating, peral2019self,omotayo2020digital,gatsou2017seniors,olson2011diffusion}) among older adults. However, it remains unknown the practices and challenges of banking on both physical and digital platforms (e.g., banking activities on each platform) among older adults and how they make trade-offs between these platforms. 

Furthermore, electronic payment services (e.g., AliPay, ApplePay, Google Pay, PayPal, Venmo, WeChat Pay, Yunshanfu, ZellePay) and virtual banks (e.g., Netbank, WeBank, Monzo) are the latest technological development in banking. Although a few studies began to explore the use of electronic payment services and virtual banks among young adults~\cite{liebana2020mobile,zhao2014exploring}, few focused on older adults. \rv{Recently, a survey study showed that electronic payment had a higher adoption rate than mobile banking and even ATMs among older adults in China ~\cite{Xiaofu21}. This has motivated us to understand the practices and challenges of using \textit{electronic payment and virtual banks} among older adults as well as other banking methods. Such an understanding may potentially explain why electronic payment, despite coming around the same time as mobile banking and virtual banks~\cite{paypal1998, MobileBanking1999, NetBank1996} and about 30 years later than ATMs~\cite{BBC2007}, has gained popularity in such a short time.} 

\rw{In this work, we seek to understand how older adults use physical and digital banking platforms, how they learn to use digital banking platforms, and the challenges encountered under current hybrid digital and physical banking services.}

\section{Method}

To answer RQs, we conducted IRB-approved semi-structured interviews with older adults to better understand their banking experiences. Moreover, we also conducted semi-structured interviews with bank employees who had experience helping older adults to gain a complementary understanding of the difficulties older adults encountered. 

\subsection{Participants}
\textbf{Older Adults}. Sixteen (N=16) participants aged 60 or older were recruited through our social network and snowball sampling. Table~\ref{tab:demographic} shows their demographic information. Nine were self-identified as female and seven as male.  They resided in different tier-ed cities: nine were in first-tier cities, five in second-tier cities, one in a third-tier city, and one in a fourth-tier city. The tier system is roughly based on the level of economic development, and the first-tier cities are considered to have the highest levels of economic development in China, such as Beijing and Shanghai~\cite{Chinesec52:online}. All participants (Md=65, SD=7) used smartphones. Five also had a computer and one also used an iPad.

\begin{table}[htb!]
    \caption{Older adults' demographic information}
    \label{tab:demographic}
    \Description{Table \label{tab:dempgraphic} demonstrates the age, sex, location, and prior device usage experiences, such as computer, android phone, iPad, and iPhone of 16 participants each.}
    \begin{tabular}{c|c|c|c|c}
    \hline
    \rowcolor[gray]{0.9}Id & Age & Sex & Location & Devices used \\
    \hline 
    P1 & 61 & F & Shenyang & computer, android phone \\
    \hline  
    P2 & 61 & M & Shenyang & computer, android phone \\
    \hline 
    P3 & 60 & M & Huainan & android phone \\
    \hline 
    P4 & 63 & M & Shenyang & android phone \\
    \hline  
    P5 & 62 & M & Jining & android phone \\
    \hline 
    P6 & 65 & F & Shanghai & android phone \\
    \hline 
    P7 & 60 & F & Zhengzhou & android phone \\
    \hline 
    P8 & 76 & M & Shenyang & android phone \\
    \hline 
    P9 & 60 & F & Shanghai & android phone \\
    \hline  
    P10 & 63 & F & Shanghai & android phone \\
    \hline 
    P11 & 60 & F & Shanghai & android phone \\
    \hline 
    P12 & 77 & M & Shenzhen & computer, iPad, iPhone  \\
    \hline 
    P13 & 79 & F & Shanghai & android phone \\
    \hline
    P14 & 71 & F & Shanghai & android phone \\
    \hline
    P15 & 62 & M & Shanghai & computer, iPhone \\
    \hline
    P16 & 60 & F & Shanghai & computer, iPhone \\
    \hline
    \end{tabular}
\end{table}

\rv{\textbf{Bank Employees}. \camera{After interviewing older adults, we found that although they wished to be assisted by their children, they tended to go to banks in person for reasons like worrying about their children not having time or relevant knowledge, not living with their children, and not wanting to reveal their financial info to their children. In contrast, they felt bank employees had worked with many older adults and had more experience helping them. Thus, we also interviewed bank employees to gain their perspectives.} Seven (N=7) bank employees who had experience helping older adults were recruited through our social network and snowball sampling. 
Table~\ref{tab:demographic} shows banking employees' demographic information. 
Five worked as \textit{lobby managers}, who were responsible for guiding and answering customer inquiries. Two worked as \textit{financial managers}, who were responsible for financial management transactions. While lobby managers could assist customers with most of the transactions, they will hand over financial management requests to financial managers. In China, lobby managers are the ones who have the most contact with customers. Generally speaking, they are the first front to serve customers and then guide customers to queue or suggest trying digital methods without waiting. They are also responsible for providing help on how to use digital banking platforms. Unlike bank tellers, who are only responsible for transactions at the counters with a fixed process, lobby managers have more diverse and extensive experience in helping older adults, which is the reason that we choose to interview lobby managers instead of bank tellers.

\begin{table}[htb!]
    \caption{Bank employees' demographic information}
    \label{tab:bankstaff}
    \Description{Table \label{tab:dempgraphic} demonstrates the age, sex, location, and job title of 7 banking staff participants each.}
    \begin{tabular}{c|c|c|c}
    \hline
     \rowcolor[gray]{0.9}Id & Sex & Location & Title \\
    \hline 
    S1 & M & Tianjin & Lobby manager \\
    \hline  
    S2 & M & Shanghai & Lobby manager \\
    \hline 
    S3 & F & Tianjin & Lobby manager \\
    \hline 
    S4 & M & Beijing & Lobby manager \\
    \hline  
    S5 & M & Shenyang & Financial manager \\
    \hline 
    S6 & M & Dalian & Lobby manager \\
    \hline 
    S7 & F & Beijing & Financial manager \\
    \hline 
    \end{tabular}
\end{table}
}

\subsection{Procedure}
We obtained approval to conduct this research from the Institutional Review Board of our university. Interviews were held through phone calls, video calls, or in-person at participants' places of choice \rw{by one researcher}. All local COVID prevention protocols were strictly followed. \camera{All participants spoke Mandarin, and we also conducted the interviews in Mandarin. While some participants voluntarily showed us their apps to explain their points, our studies did not ask for or record any personally identifiable information.}
The interviews lasted about an hour \rw{ranging from 47 minutes to 76 minutes} and were audio-recorded. Each participant was compensated with 70 CNY. 

\rv{For older adult participants,} they were first asked to share their banking practices including the activities they performed on various banking platforms (e.g., physical banks, ATMs, mobile banking apps, virtual banks, and electronic payments) and the reasons for their choices. Following a semi-structured format, we asked them to freely talk about their usage patterns among these platforms and the challenges they encountered as well as the strategies they used to overcome the challenges. Specifically, we asked participants to explain their problems by showing the issues with their apps for both in-person and video-call interviews. For phone-call-only interviews, we installed the same app as the participants to follow their descriptions of the problems. Lastly, we asked about their expectations of banking platforms. 

\rv{For banking employee participants, they were first asked to estimate the proportion of older adult customers and their queuing status(e.g., whether they need to queue, how long they line up), and to share their observations and experiences regarding older adults' practices and challenges of using physical counters and digital banking methods. They were also asked to talk about the types of transactions that they often help older adult customers with. After that, they were asked to share the difficulties they encountered when helping older adult customers and their banks' policies for older adults. Finally, they were asked to share how they felt banking systems could be improved to better help older adults.}

\subsection{Data Analysis}
\rw{The interview recordings were auto transcribed, and one native Mandarin-speaking author reviewed and corrected the transcripts.} Two Mandarin-speaking authors first familiarized themselves with the transcripts and then coded them independently using an open coding method~\cite{corbin2014basics}. They met to discuss their codes and rationals, revise their codes, and resolve disagreements to gain consensus on the codes. After that, all the codes were translated into English, and all the researchers performed affinity diagramming~\cite{Beyer1997Contextual} to \rw{cluster} the codes and \rw{categorize} emerging themes \rw{related to our research questions with an inductive approach}. We followed an iterative process to constantly challenge the groupings until we reached a consensus on the final groupings and their themes. 
\section{Findings}
Our analysis revealed three main themes about practices and challenges of banking among older adults: 1) Activities and Practices on Different Banking Platforms, 2) Motivations to Learning Digital Banking Platforms, and 3) Challenges in Using Digital Banking Platforms. We refer to older adult participants as ``participants'' and bank employee participants as ``bank employees'' in the rest of the section.

\subsection{Activities and Practices on Different Banking Platforms}


\subsubsection{Banking Transactions on Physical and Digital Platforms}
\label{sec:practices_banking}
\begin{figure*}[htb!]
  \centering
  \includegraphics[width=0.99\linewidth]{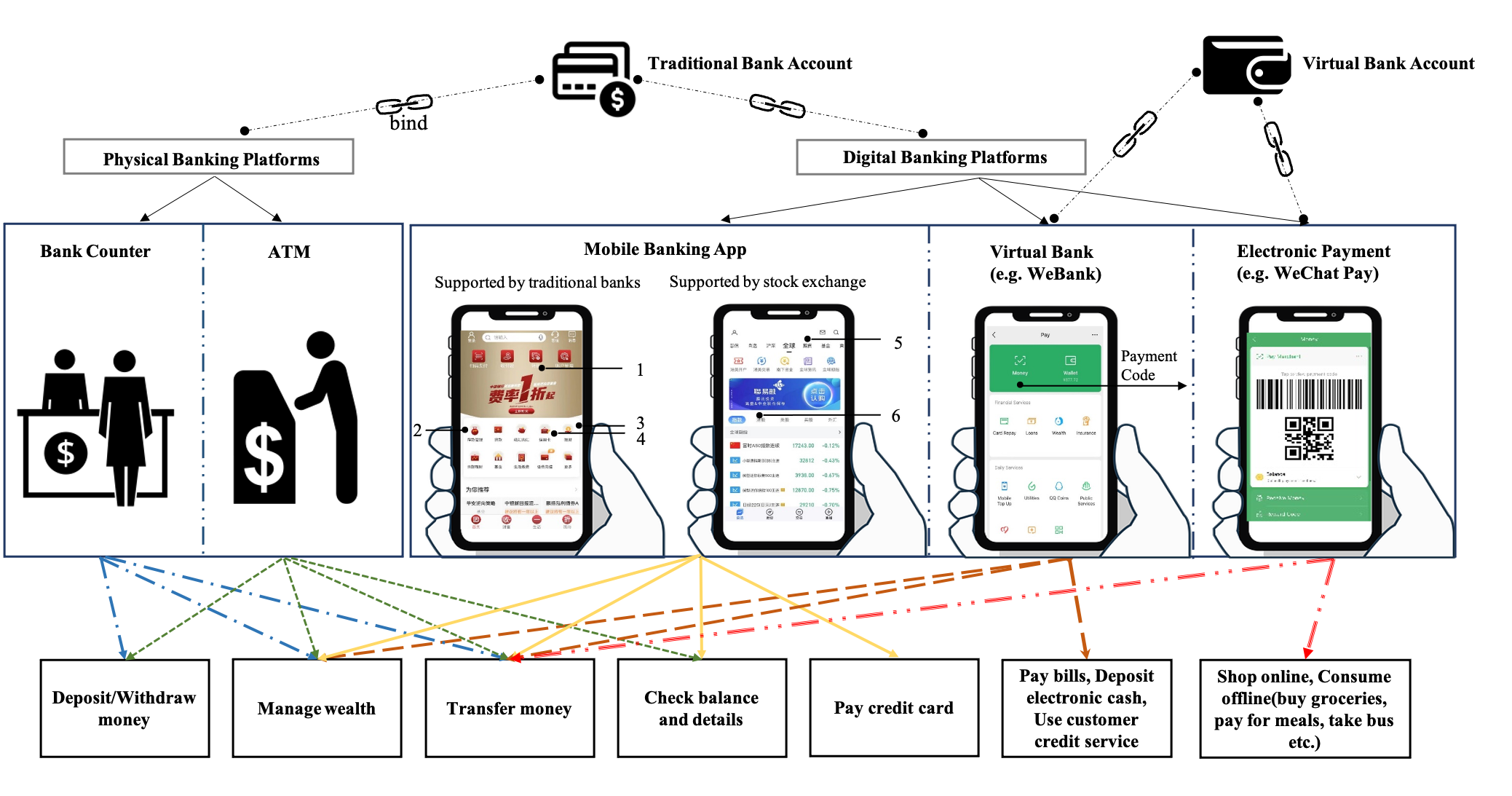}
  \caption{Participants' banking platforms and transactions. An arrow from ``traditional bank accoun'' points to both physical banking platforms and digital banking platforms, which indicates that these platforms could be bound with traditional bank accounts. Another arrow from ``virtual bank accoun'' points to virtual banks and electronic payments, which indicates that these services could also be bounded with a virtual bank account. Different colors and types of the lines indicate the connections between the banking platforms and the corresponding transactions that older adult participants performed.}
  \Description{Figure \label{fig:PlatformDescription} shows the connections between bank accounts, banking platforms, and bank transactions. All components are connected to form a tree-like structure. At the root level, there are two items: "traditional bank account" and "virtual bank account." At the second level, there are two categories of banking platforms: physical banking platforms and digital banking platforms. At the third level, there are specific banking platforms in each category as the text talks about. At the fourth level, there are common banking transactions older adults conduct.}
  \label{fig:PlatformDescription}
\end{figure*}

Figure ~\ref{fig:PlatformDescription} shows participants' banking platforms and transactions. 
Physical banking platforms include both bank counters (i.e., physical banks) and ATMs. Digital banking platforms include mobile banking apps, virtual banks, and electronic payment. Next, we present the details of the transactions on each platform. 

\textbf{Physical Platforms}.
Participants primarily conducted the following transactions in physical bank counters: deposit money, withdraw money, manage wealth, and transfer money. Likewise, participants also conducted a similar set of transactions on ATMs though they tended to use ATMs more often to check balances and details of their bank accounts.

\rv{Bank employees estimated that older adult customers took up 30\% to 90\% of all the customers visiting physical banks, depending on the bank branch's location, the date (e.g., day of pension release), and the time of day. \camera{Our analysis found that the wide range ([30\%-90\%]) was mainly related to banks’ locations and types. Banks on the higher end of the range tended to be state-owned big banks with a long trustworthy history and locate in residential areas. The ones on the lower end tended to be relatively small banks or locate in business districts. Moreover, bank employees reported that older adult customers tended to flood to the banks around the time when pension payments are released, which crowded the banks and overburdened the bank employees. Furthermore, bank employees reported that older adults who did not use smartphones faced extra challenges during the COVID-19 pandemic because banks required their customers to scan the QR code to reveal their health code on their phones to prove they are healthy before being allowed to enter the banks, which were complained a lot by them.}}

\textbf{Digital Platforms}.
Participants used three main forms of digital banking platforms: mobile banking apps, virtual banks, and electronic payment. 
Participants used two types of mobile banking apps to conduct financial transactions remotely. One type was provided by traditional banks (see the left UI in the middle of Fig.~\ref{fig:PlatformDescription}), and they used this type of mobile app to transfer money, manage wealth, and conduct transactions related to a credit card. \rv{Bank employees confirmed that mobile banking apps provided by traditional banks provided the majority of the non-cash banking transactions.}

The other type was provided by or in collaboration with \rv{stock exchange} (see the right UI in the middle of Fig.~\ref{fig:PlatformDescription}), such as Eastmoney and Dazhihui \cite{Eastmoney, Dazhihui}, and participants used such apps mostly to perform wealth management, such as buying stocks or funds.

Virtual (online-only) banks are the ones that do not have physical branches but offer banking services remotely \cite{Wiki-VirtualBanking}. Participants used virtual banks' apps to deposit money to their electronic payment account (e.g., WeChat Pay~\cite{Wechatpay}---an electronic payment app), pay bills, and manage wealth. 
Furthermore, participants also used the virtual bank as a bridge to transfer money between accounts in different physical banks. They would first transfer money from one bank account into the virtual bank and then transfer the money again out to another bank account. In such a way, they could avoid a transaction fee associated with transferring money directly between two bank accounts because virtual banks have the benefit of waiving online transferring transaction fees.

Electronic payment options include digital wallets (e.g., Alipay Wallet, WeChat Wallet), contactless payment methods (e.g., Alipay, WeChat Pay), etc. To use electronic payment, participants needed to create a virtual account and bind one bank account with the virtual account so that they could transfer money from their bank account into the virtual account. The two most commonly used electronic payment services among participants were WeChat Pay \cite{Wechatpay} (Fig.~\ref{fig:PlatformDescription} right shows two UIs) and Alipay \cite{Alipay}, and participants used them regularly for daily activities, such as buying groceries, shopping online, paying for meals, taking bus or taxi, and transferring money.

\subsubsection{Perceived Pros and Cons of Banking Platforms}
\label{sec:pros_and_cons}

\textbf{Physical bank counters}.
\rv{Safety was the major perceived pros of physical bank counters. 
In addition to a safe physical environment, being able to correct errors with the assistance of bank employees was another reason for perceived safety. \textit{``I would run into big trouble if I accidentally typed 10 instead of 5 years for the deposit time on my phone if I'm alone. But I could correct it right away with bank employees.}''- P3.}

On the other hand, the long wait time was a frequently mentioned disadvantage of physical bank counters. 
\rv{Bank employees reported that wait time was influenced by the bank's location, date, and the time of a day. For example, wait time would be longer if a bank branch was the only one in the local area, or if it was the pension release day. Especially on the day of pension release, sometimes even before the bank branch is open, people have already filed a long line at the bank gate. Moreover, they mentioned that to comply with improving regulatory requirements, the transaction processes became longer than before, which further increased wait time.}

\textbf{ATMs}. ATMs were perceived to be convenient and time-saving for certain banking transactions, such as withdrawing money. 
However, participants complained that they were unable to find out whether ATMs had enough cash until they inserted their cards and were several steps into the process. 
What's more, participants were concerned about the legibility of the ATM's screen and often had to deliberately bring reading glasses to overcome this problem. ``\textit{Who would always remember to bring the reading glass when going out?}''- P1. \rv{Indeed, bank employees also shared similar feedback from older adult customers and pointed out some bank branches started to provide reading glasses to customers.}
Moreover, participants were also concerned about the fact that others could spot their passwords over their shoulders when they used ATMs. 

\rv{Furthermore, bank employees mentioned that the latest versions of ATMs allowed customers to sign on the screen using a digital pen. However, many older adults had hand tremors, which made it difficult for them to sign on the screen. }

\textbf{Digital Platforms}. 
One common advantage of all digital banking platforms was convenience. First, with digital platforms, participants could save physical efforts while still satisfying their banking needs at home. ``\textit{I could pay utility bills using AliPay (a virtual bank) with a few clicks and don't have to go out}''-P2; ``\textit{Those unnecessary human labors or human errors are saved by the information technology.}''-P5 \rv{Second, participants did not need to carry and deal with banknotes, which were susceptible to both hygiene and counterfeit issues. In particular, participants appreciated that they did not have to carry their purses or wallet to store cash but just needed to carry their smartphones, which almost all of them would carry anyway when going out. 
\textit{``I used to carry a wallet and need another bigger bag to store it along with my phone. Now, I only need to carry a small bag to store my phone.''}-P6; \textit{``It helps to avoid touching paper bills and is more hygienic; this is especially important during the pandemic.''}-P5. Third, participants could transfer the money instantaneously on digital platforms. P5 preferred to use \textit{pocket money} (i.e., a function of WeChat Pay~\cite{Wechatpay}) to transfer money to his daughter, who studied far away, over traditional money transfers in banks because the money would arrive at his daughter instantaneously to ensure her financial needs without delay.   }

On the other hand, participants expressed privacy and security concerns about digital banking platforms. 
\h{Because of the security concern about the electronic payment (e.g., WeChat Pay), P7 did not directly bind her bank account with WeChat Pay. Instead, she usually withdrew more cash from physical banks than she needed and exchanged the extra cash with her trusted relatives so that they could transfer money to her WeChat Pay. }
Moreover, \h{participants also had security concerns about offline transactions. When paying a person offline (i.e., in-person) using an electronic payment app, the app typically generates a barcode and displays it on their phone's screen for the person to scan. However, participants worried that passers-by might be able to scan the barcode easily from a distance and steal their money. \textit{``A Passer-by could easily use their phone to scan the barcode on my phone from several meters away without my awareness, so I never use it offline.''}-P7 }
\subsection{Motivations to Learning Digital Banking Platforms}


We identified three primary motivations for learning digital banking platforms: perceived usefulness, self-confidence, and social influence. 

\textbf{Perceived Usefulness}. Participants felt that it was convenient to complete transactions for daily activities, such as paying groceries and transportation fees, on their smartphones, especially for daily activities. \camera{Bank employees also reported that many older adults were willing to try digital banking platforms even if they had to wait in lines for a long time (e.g., over an hour).}

Moreover, they were also motivated by economic benefits, such as discount activity on online shopping platforms.
\textit{``After hearing from others that fruits on Pinduoduo (an e-commerce platform) were cheap, I started to learn to use WeChat Pay.''}-P16. Similarly, they were also motivated by the fact that many offline shops offered discounts when customers pay with electronic payment. \textit{``Yunshanfu (an electronic payment app) collaborates with local shops to offer discounts on their groceries, such as tofu, so I was motivated to learn to use it.''}-P8. Additionally, one participant was motivated by higher interest rates of wealth management products offered through mobile banking apps. \textit{``Some wealth management products have higher interests, but they are only available in mobile banking apps, so it is worth learning.''}-P14.

Another perceived usefulness of learning digital banking platforms was to keep their brains sharp. \textit{``I'm getting old, and if I don't use it, I'll lose it.''}-P14.



\textbf{Self-Confidence}. Overall, participants expressed confidence in their ability to learn digital banking. \textit{``I might be slower than others, but just because I'm slow doesn't mean I can't learn it.''}-P6; \textit{``Although I am slower than young people, I could still manage to learn it.''}-P14; 

Moreover, their prior experiences with technology seemed to help them gain confidence in learning digital banking platforms. P12 was used to learning things by himself because he did not live with his children for a long time.  ``\textit{Many years ago I bought a computer and started to learn to type, and I got used to this learning process over time, so it's not too hard for me to learn these operations on my phone.}''-P12.
Similarly, P15 learned how to do wealth management on the website using a computer and felt it was relatively easy to transit conducting on mobile banking apps.

Besides, they tended to describe learning digital banking as troublesome instead of difficult. \textit{``It was troublesome to use mobile banking for wealth management because I only did it once per year but still had to remember the steps.''}-P4. Last but not least, they were able to afford the learning efforts in particular if they could receive some help along the way. \textit{``Someone taught me how to use it, and I felt it was not that hard to learn.''}-P1.

\textbf{Social Influence}. All participants mentioned that they were motivated by people in their social circles, such as their children, grandchildren, and friends, who almost all used digital banking platforms. 
\h{P15 mentioned that he decided to learn to use digital banking because he envied his colleagues for being able to use digital banking apps to buy breakfast much faster. \textit{``I was reluctant and resistant to learn it, but later I noticed that my younger colleagues used their phones to pay for breakfast by simply scanning a code. In contrast, I had to carry change to pay for my breakfast, which was much slower. I really envied them and wanted to learn. Later, they taught me how to use it.''}-P15.}
\rv{Bank employees also reported that there are few older adults who would like to proactively propose to use digital banking methods to replace manual counters although most of the banking transactions could be conducted via digital banking platforms. Therefore, bank employees provide active guidance and advertisement to encourage and lead older adults to use digital banking tools like mobile banking apps. While there were still a large proportion of older adults who would rather wait a long time for the manual counters, which consist with the findings of the study~\cite{Xiaofu21}. Some older adults were indeed motivated by their words and tried these new methods with their guidance.}

\h{Moreover, participants were motivated to learn because they were involuntarily connected with digital banking platforms through their social connections. P16 started to learn digital banking because her friend transferred digital cash to her when paying her back. She thought that she had to find a way to spend the money in the digital bank, so she gradually learned to use electronic payment and now she could use it for online shopping.}

Furthermore, participants felt that digital banking represented a societal trend and they needed to keep up with the fast-changing world and not become obsolete. \textit{``Society is moving forward, and new technology is beneficial for societal development.''}-P5; \textit{``Now everything is online, so if you don't know how to do it online, you will be outdated.''}-P6.


\h{Lastly, participants also felt proud of being able to use digital banking among their peers. P8 and P14 regarded being able to use digital banking \textit{as a fashion}. \textit{``I think learning digital banking is a kind of pride and fashionable.''}-P8.}

\subsection{Challenges in Using Digital Banking Platforms}
\label{sec:challenges}
When using digital banking platforms, participants encountered app-related challenges and user-related challenges.

\subsubsection{App-Related Challenges}
 As for app-related challenges, P6's complaint may shout out many older adults' thoughts, \textit{``It is not designed for our older adults, it is designed for the young``-P6}. Besides participants, bank employees confirmed that many older adult customers complained about the \textit{poor design and defective accessibility} of the banking app. There were four types of app-related challenges: ambiguous affordance, low information scent, insufficient error recovery support, and lack of feedback or confirmation. 


\begin{figure}[htb!]
  \centering
  \includegraphics[width=0.8\linewidth]{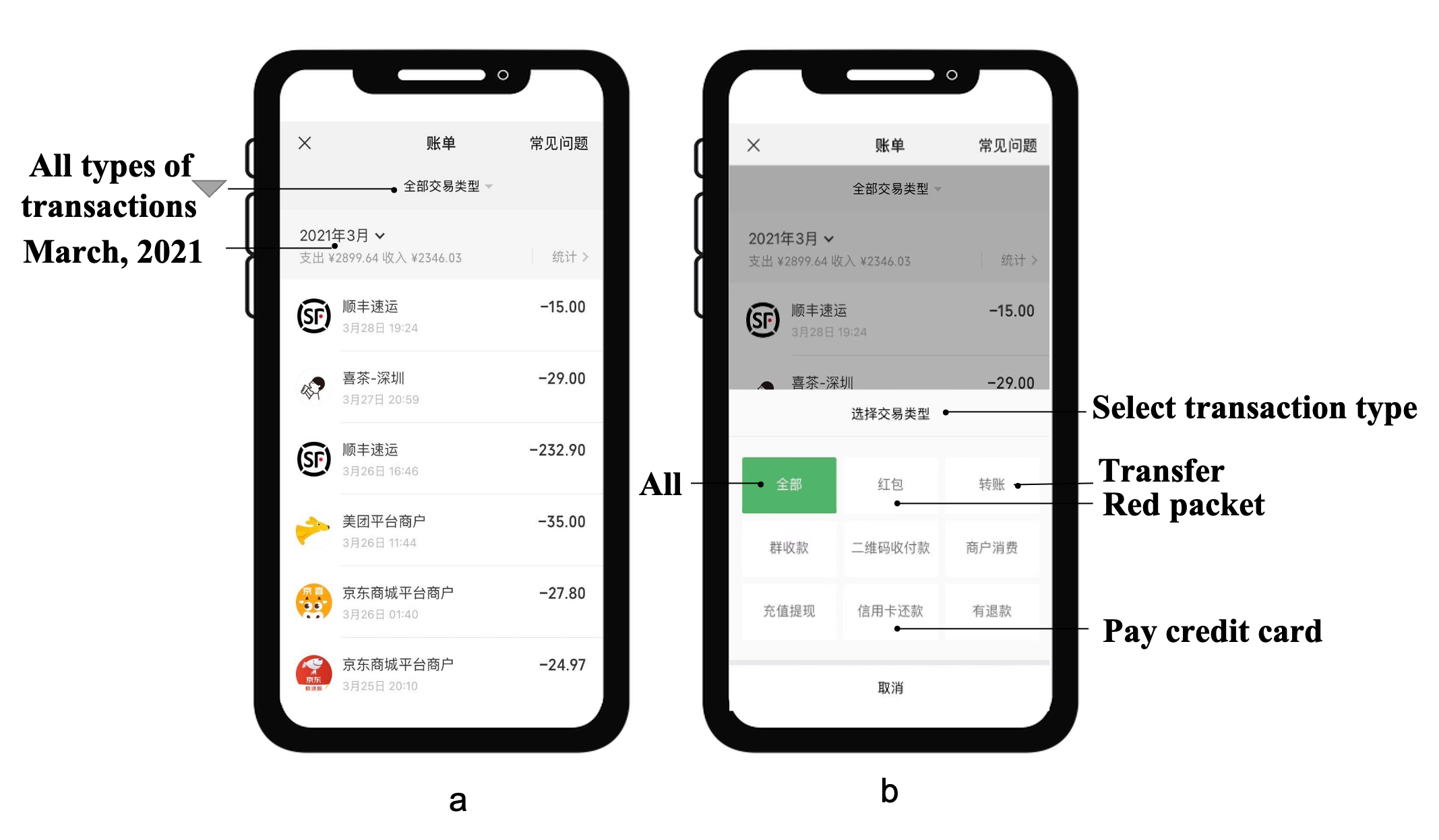}
  \caption{UIs for checking bills in a virtual bank (WeBank~\cite{WeBank}): (a) a clickable option besides ``All types of transactions''; once clicked, a pop-up window shows up; (b) The pop-up windows allows users to select and view the transactions of a specific type.}
  \Description{two screenshots that show (a) a page of checking bills, (b) a popping up window of (a). At the top of screenshot (a), there is a button labeled ``All types of transactions" with a clickable option nearby. Beneath it, there is a list of records of bills. In screenshot (b), the popped-up window takes half of the page at the bottom. It shows nine options that could be clicked on for choosing the category the user would like to show up.}
  \label{fig:bills}
\end{figure}

\textbf{Ambiguous Affordance}.
Participants did not realize certain functions existed due to the ambiguous affordance of UI elements. For example, Fig.~\ref{fig:bills} a shows a list of ``all types of transactions'' in March 2021. There was a gray triangle icon on the right side of the title. When clicked, the interface pops up a window (Fig.~\ref{fig:bills} b), which includes an ``All'' option for users to check their spending on all types of transactions and other options, such as \camera{``Red packet''}, ``Transfer'', ``Pay credit card,'' for them to check the spending of a specific type. However, the triangle icon was not perceived as clickable. As a result, P1 believed that there was no way to check her total spending: \textit{``WeBank shows all transactions together. When I had too many transactions, I needed to check my spending on a specific category. But I could not do it.''}

\textbf{Low Information Scent}.
One symptom of low information scent was that participants had no idea certain functions existed in the app.
P5 was frustrated with not receiving real-time notifications about his spending when using the credit function of a virtual bank. When the credit bill was due after a month, he often found he spent more money than he realized: \textit{``It wasn't transparent and didn't send any notification in time. I only realized the bill was higher than I expected after a month.''}
This confusion was caused because P5 did not realize that the real-time notification needed to be enabled in the settings. 



Moreover, participants were also frustrated about being redirected from page to page. \textit{``ATM has clear steps, but the app does not. You have to click here to enter a page, then click there to go to another page, and so on and so forth.''}-P1. Bank employees reflected that it did happen commonly during the use of banking apps and it was a kind of burden for them to lead older adults to the exact page as well.

\textbf{Insufficient Error Recovery Support}.
The digital banking apps did not provide sufficient guidance for participants to recover from a mistaken operation. \textit{``I once clicked on a wrong thing and was taken to a completely different place. I had no idea how to return to the previous step and became panicked. In other situations, I would have restarted my phone, but this was related to money and I was afraid of losing money if I restarted my phone.''}-P6. Thus, P6 went to the bank to ask for help instead, but she felt this approach was too costly to be a regular solution.



Moreover, insufficient error recovery support could also happen at the operating system level. P8 reported that he once could not find the electronic payment app after he accidentally switched to other apps with an unintended touch gesture. \textit{``I didn't know how I switched to a different app [from the WeChat Pay app]. I tried to swipe in different directions to get it back but I couldn't.''}-P8

Actually, participants expressed that they would like to do trial and error by themselves because they could learn without troubling others. Moreover, they felt that learning independently would help them gain a deeper understanding and remember it longer. \textit{``If I learn and do it myself, I would remember the process much better. My wife always asked our kid to teach her and she still couldn't remember the steps. Thus, getting help from others is no better than learning it by me.''}-P2. After receiving a text message with a link from his bank, P2 clicked the link on his phone, downloaded the banking app, and tried to bind his bank card to it. However, without sufficient error recovery support, he felt reluctant to try those unfamiliar functions in the app.

\textbf{Lack of Feedback or Confirmation}.
Participants also mentioned that their apps did not provide confirmation or feedback when they finished typing their bank account number and submitting it. Consequently, they were unsure if they were successful or not and therefore were hesitant to continue the following steps. \textit{``The app has no confirmation after I enter the amount to be transferred. As a result, I have to check many times to make sure I have entered the correct information. Because of this, I am afraid of transferring any large amount.''}-P9




\subsubsection{User-Related Challenges}
\camera{Participants encountered the following user-related challenges: digital banking's intangibility, trustworthiness, security and privacy, overspending concerns, low perceived self-efficacy, and memory concerns.}


\textbf{Intangibility.} Unlike physical banks that had a physical location and could provide printed receipts for transactions, digital banks were perceived to be intangible. ``\textit{I could not see what I buy in virtual banks.''}-P6. However, the worry about abstraction seemed to go away as they gained more experience with digital banking over time. ``\textit{I was hesitant to use digital banks in the beginning, but I gradually accepted it after using it for a while.}''-P3.
 


\textbf{Trustworthiness.} The trustworthiness of digital banking platforms also made participants worry. Unlike physical banks, in particular national banks (state-owned capital holding banks), that had been there in their entire lives, digital banks were new. They tended to think of digital banks the same as small local banks and worried about their trustworthiness. They mentioned stories about the bankruptcy of small banks. \textit{``Small local banks (whose the controlling shareholder is the local government or private capital) may run into cash flow issues and could even bankrupt like Haifa bank, so although I use digital banks, I still worry about not being able to get my money out.''}-P5.
\rv{Participants' doubts in small banks were reasonable. None of the nation-wise banks have bankrupted in China. Although rare, one local bank (Hainan Development Bank) did bankrupt in 1998~\cite{Haifa}. 
What’s more, digital platforms tend to be riskier than physical ones. For example, hundreds of digital peer-to-peer lending platforms were found to be fraudulent in 2018~\cite{P2P}. The rampant fraud targeting older adults also made older adults more cautious and anxious about digital banking.}



\textbf{Security and Privacy.} The security and privacy of digital banking also caused anxiety. Participants mentioned that they heard about online hackers and data breaches in the news. 
Moreover, they also worried about losing money due to mistakes in electronic payment. \textit{``What if there was a mistake in electronic payment and others swiped my card?''}-P3. 
Lastly, unlike physical banks that offer printed receipts, they felt virtual banks did not provide receipts that they could use to go and get the money back if unexpected things happened. \textit{``I don't have any physical receipts. What if the digital bank disappears all of a sudden like the app is gone, what do I do with my money?''}-P6. \rv{Bank employees echoed that some older adults even did not trust the receipts printed by ATMs and only trusted the ones from a bank counter.}

Participants also mentioned privacy breach though they were unsure whether it was due to physical or digital banks. \textit{``I received a lot of spam calls from all over the country. I didn't know who leaked my phone number but I suspected it was the banks since many calls are for loaning stuff.''}-P2


\textbf{Overspending Concerns.} Participants felt that it was harder to keep track of their spending when using digital banking and consequently they tended to overspend. they. \textit{``Sometimes I was surprised how much I had spent when I was reviewing my monthly bill.''}-P6; \textit{``It (Digital banking) was indeed convenient. But I ran out of budget much more easily [using it] than using cash. With cash bills, I was prompted to think again about how much I would spend.''}-P8


\textbf{Low Perceived Self-Efficacy}.
Another user-related issue was caused by low perceived self-efficacy. One recurring symptom was being afraid of making mistakes. This was because making mistakes would likely cause them to lose money. \textit{``When buying stocks in mobile apps, I am extremely worried about making mistakes. For example, it is very easy to type an additional zero by accident.''}-P1 
The concern about making mistakes was even severer when the mistakes were perceived as nonrecoverable. 
Reasons for low perceived self-efficacy included the lack of confidence in their literacy level and their declining physical conditions, such as poor eyesight. 
\textit{``I don't have much education, and it is common for me to press the wrong button.''}-P3; \textit{``I recommended WeChat Pay to my sister. However, she resisted it because she was worried about making mistakes and losing money due to her poor eyesight.''}-P2
In this case, they preferred to learn to use digital banking with assistance from their close peers (e.g., colleagues, classmates, friends), their family members (e.g., children, relatives), banking employees, and classes offered by local communities and volunteers, that made they felt reassurance about learning. 

When deciding from whom to seek help, participants reported considering both the trustworthiness and expertise of the help givers. 
\rv{Regarding trustworthiness, participants preferred to be assisted by people who knew them well. \textit{``My children know my personality and knowledge level. Thus, they would teach me at my level. But others might start from a higher level. They may be kind and helpful, but I won't be able to understand it.''}-P8. Moreover, participants also worried that help givers might lose patience and assist them less carefully, especially when there were other people waiting to get help from the same help giver. 
Indeed, bank employee participants also reported that in peak hours they could only tell customers where to click with little time for an explanation so as not to keep other customers waiting for too long. In some cases, they would even operate directly on older adults' phones instead of guiding them even though this was discouraged by bank regulations for privacy concerns. \textit{``The pressure of serving all customers in line is high, so it is challenging to be always patient and meticulously. In fact, more often than not, the regulation of never operating directly on customers' devices was not well followed.''}-S6. From older adults' point of view, many were willing to let bank employees operate on their phones, except for inputting the password, to save time and effort. \textit{``Many older adults mentioned that they had a bad memory and might forget it easily after being taught, so it was better to operate it for them directly.''}-S1.}

Furthermore, some participants preferred help givers to monitor their operation process and only give feedback when they ran into trouble. ``\textit{I'd like the bank employee to stand nearby to monitor my progress while I'm trying to do it on my phone. I hope that he would help me only when I run into trouble. Otherwise, I won't be able to do it next time." -P6}. \camera{Bank employee participants confirmed that many older adult customers wished to have them oversee their operations to spot mistakes and provide answers, such as ``\textit{I am going to tap on this button. Are you sure I tap on the right button?}'' -S2. Moreover, bank employees reported that many older adults had the ability to conduct transactions in banking apps by themselves, ``\textit{The majority of them (older adults) did the right operations all the time, but they just wanted me there to confirm each operation.}''-S2.}

Eventually, they still hoped to become as independent as possible to avoid troubling others. \textit{``Young people are busy with their jobs. I try to avoid bothering them and learn by myself''}-P6. To help themselves become independent, participants felt it helpful to have a voice assistant whom they could ask for help. \textit{``It would be very helpful to have a voice assistant. Whenever I didn't understand it, I could just ask for help. I don't need to go to banks to ask their employees. I don't want to bother others.''}-P7. \camera{Meanwhile, bank employees expressed their concerns that they might get impatient during busy hours with older adults asking repetitive questions, as S6 reflected, ``\textit{If some older adult customers keep asking me the same question while there are so many people waiting in lines, my voice may sound impatient}''. This might also explain why older adults feel bad about troubling others and want to be as independent as possible.}

Besides learning with others' assistance, during the usage of digital banking platforms, participants developed four strategies to manage the risk of losing money. First, they set an upper limit for the amount of money that they would feel comfortable with even if they would lose it. \textit{``I only put in 1000 CNY. Even if I became a victim of a fraud, I would only have that much to lose, which is manageable for me.''}-P3.  
Second, they would not directly bind their regular bank accounts with the digital platforms. Instead, they bind a dedicated non-primary bank account with digital banking platforms. In this way, they could avoid the risk of losing money in their primary bank accounts. 
The third strategy was to practice and get familiar with digital banking platforms with a small amount of money first. After several successful trials, they would gradually increase the amount. For example, when P6 started to use her virtual bank, she was concerned about whether it was easy to withdraw money from it. After successfully withdrawing money several times, she became more confident about it and started to put more money into it.  \textit{``I just wanted to withdraw money from AliPay (a virtual bank) to test if it worked. After realizing it worked well, I would deposit the money back.''}-P6. Nonetheless, participants still tended to set a maximum amount for digital platforms.  
Lastly, they believed that ``avoiding petty discounts'' could also help them avoid falling for online frauds. 

\textbf{Memory Concerns}.
Another category of the user-related issue was about forgetting various types of information related to digital banking. 
First, they were concerned about forgetting the steps of completing a bank transaction. \textit{``I followed the bank employee's instructions to walk through the process step by step. However, if I didn't write it in my notebook, I would forget about the steps.''}-P11

In addition, they also worried about not being able to remember how to use many functions at the same time. Two of them mentioned that after mastering basic functions, such as how to use electronic payment to pay bills, they worried that they might forget how to use basic functions if they continued to learn more advanced functions. \textit{``I can't put too much information in my head. Otherwise, I would mess it up and forget about it altogether.''}-P13 


Second, they were also concerned about forgetting passwords.
P11 reported that she forgot the password and did not know how to find it back. The mobile banking app asked her to set the password to be a combination of numbers and alphabets to ensure its security. However, such requirements increased the burden of memorization for older adults. For example, after many failed attempts, P11 finally had to work with a bank employee to get her money back and never used the mobile banking app afterward.

\section{Discussion}

Prior research investigated how older adults use physical banks~\cite{abood2015can}, ATMs~\cite{Nedaei2013}, and online banks~\cite{omotayo2020digital}. 
Although some researchers studied how older adults use different banking platforms, they primarily offered a quantitative overview of their usage via surveys~\cite{mattila2003internet,olson2011diffusion,Xiaofu21}.
Moreover, electronic payment services (e.g., AliPay, ApplePay, and WeChat Pay) have been increasingly integrated into our societies in recent years. In particular, China has the highest growth rate of electronic transactions~\cite{The2020M23:online} \rv{and becomes the current world leader in the usage of proximity mobile payments, among which 81.1\% of smartphone users adopt proximity mobile payments~\cite{EMarketer2019Mobilepayment}}.
Although a few recent studies explored how electronic payment services were used, they mainly focused on young adults\rv{~\cite{liebana2020mobile,Lin2020Exploring,zhao2014exploring}}. While a recent study provided a quantitative understanding of how older adults use electronic payment through a survey study~\cite{Xiaofu21}, it did not explain why older adults chose to use it over other banking methods. 
Building on top of prior work, our work presents an in-depth qualitative understanding of why and how older adults use both physical and digital banking platforms in today's technology landscape, how they learned to use digital banking platforms, and the challenges that they encounter in a country that has been experiencing one of the fastest growth of electronic transactions. \rv{Furthermore, our work also provides complementary perspectives from bank employees to better understand older adults' banking practices and challenges.} We discuss our key findings, the design implications, and potential future work in this section.
\subsection{Banking Activities and Practices}
\rv{Front-line bank employees reported that almost all older adults came to the banks by themselves, showing their intention to be independent. Older adults also had their own usage patterns with considerations of different platforms' pros and cons. Although their understandings of different platforms varied in some aspects, they generally agreed that \textit{security} and \textit{timely in-person assistance} are two advantages of physical banks with a series of visible secure checks and professional staff. One the other hand, the complexity of the secure process combined with other factors like the bank branch's location and the date (e.g., the pension releasing day) also caused inconvenience, such as the long wait time in physical banks. In contrast, using digital banking platforms could mitigate this issue and bring \textit{convenience} for either conducting banking transactions online or paying offline without cash. However, it remains an open question of how to design a hybrid online and offline ecosystem to integrate the advantages of the two. 

Bank employees mentioned that almost all the banking transactions can be conducted through digital banking platforms except the cash-related ones. Prior studies pointed out the two most common transactions conducted by older adults were \textit{money transfers} and \textit{account/transaction inquiries}
~\cite{Xiaofu21,omotayo2020digital,camilleri2017relevance}. This is consistent with what we observed in our study. Participants frequently mentioned their experiences of transferring money and checking balances or details. They also reflected challenges related to these two types of transactions (Sec.~\ref{sec:challenges}), which suggests that even the most common transactions still need to be improved. Future work could investigate ways to  improve the accessibility of older adults' commonly used transactions as a stepstone to increase their adoption of digital banking platforms. 

Furthermore, both older adults and banking employees summarized the characteristics of banking transactions that older adults conducted as being relatively simple and having a relatively low risk. Reasons for not conducting more complex and risky transactions among our participants include the concerns of losing money and the lack of financial knowledge. Thus, future work should investigate ways to help older adults confidently learn to perform more complex transactions.}

\subsection{Potential Cultural Effects}
\camera{Our study conducted with older adults living in China, and their banking practices may have been influenced by the Chinese culture that differs from those of western countries. Our participants reported some practices that were not found in western countries. For example, older adults in China widely use digital payment functions in WeChat, a social media app, to pay for daily expenses (e.g., groceries, transportation). In contrast, we have not found a study reporting a similar trend in western contexts. During Chinese traditional festivals, older generations often gift younger generations a physical “red packet”, which is a red envelope with cash in it. Interestingly, WeChat allows them to continue practicing this tradition by sending a digital “red packet” with a similar visual effect to each other conveniently.

On the other hand, there are common practices among older adults in western countries, which are uncommon in China. For example, older adults in western countries (e.g., the UK, the US, Spain, Canada) regularly use personal cheques~\cite{Vines2012Questionable, Latulipe2022Unofficial, abood2015can}. However, people in China in general rarely use them. Moreover, while it is common for older adults in western countries to use credit cards, many major banks in China impose age restrictions on credit card applications and exclude older adults~\cite{Wang2015OlderAdultsCreditCard}. This restriction might explain why older adults in China have less experience with credit cards~\cite{Xiaofu21}. One reason for rejecting older adult applicants is that banks lack confidence in their ability to pay credit bills~\cite{Wang2015OlderAdultsCreditCard}. However, such a concern was not reported for western contexts in prior studies. 
Furthermore, older adults in Canada were reported to be comfortable with sharing online banking credentials with their close ties~\cite{Latulipe2022Unofficial}. In contrast, our participants tended not to want to reveal their credentials to their close ties, who would then know their financial status. Instead, they were comfortable sharing their credentials with bank employees when seeking help because they felt that bank employees’ work ethics would not allow them to abuse their credentials. In sum, although our research scratched the surface of the potential effects of culture on older adults' banking practices and challenges, more systematic and substantial follow-up research is needed to further investigate the cultural effects.
}

\subsection{Motivations to Learn Digital Banking}
We found that older adults learned digital banking with three primary motivations: perceived usefulness, social influence, and self-confidence. \camera{These three motivations relate to several dimensions in the UTAUT2 (Unified Theory of Acceptance and Use of Technology) model~\cite{Venkatesh2012UTAUT2}. Specifically, the ``perceived usefulnes'' relates to both ``expectation of performance'' and ``price value'' in UTAUT2, the ``social influence'' relates to ``social influence'', and the ``self-confidence'' relates to``expectation of effort''.}
\rw{Unlike the motivation of learning social technologies for communication such as reducing the loneliness, which happens more prevalent in older adults~\cite{nicolaisen2014lonely,sims2017information}, participants learned digital banking because they felt that digital banking was convenient and offered economic benefits. \rv{These motivations aligned with younger adults' acceptance of online banking and electronic payments, which included ``perceived ease of use, perceived values, and perception of utility''~\cite{Hua2008, liebana2020mobile}.} In addition, older adults felt learning digital banking could keep their brains sharp. This echoes the finding of the previous study that older adults play digital games to help treat age-related cognitive disorders~\cite{Cota2015Mobile} and learn computer programming to keep their brains challenged~\cite{Philip2017Older}. \rv{Compared to the literature, the motivation to learn new technologies such as digital banking platforms for keeping brains sharp is perhaps a unique motivation among older adults, while young adults rarely mention it as a motivation.} 

Social influence was found to affect younger adults' adoption and continued use of mobile payment services recently~\cite{Lin2020Exploring}. Our work extends this finding and confirms that it also affects older adults' adoption of digital banking platforms. Participants were motivated by the adoption of digital banking among people in their close social circles and felt they should learn it to keep up with the social norms. This motivation echos the previous research that older adults are motivated and encouraged by people around them such as their children and grandchildren to learn to use mobile phones and computer devices ~\cite{luijkx2015grandma} and do online shopping  ~\cite{Lian2014Online}. 

However, in contrast, previous research showed that, unlike other ICTs, the intention to learn internet banking among older adults is not significantly impacted by social influence since users tend to access banking services alone considering security issues~\cite{yuen2010internet,arenas2015elderly}. One potential reason for this difference might be that the new generation of digital banking services offers more social opportunities, such as sending digital money (e.g., a red \camera{packet}) to their social circles as financial and emotional support or seeing others pay for goods with a single contactless tap, that was unavailable decades ago. 
As our participants mentioned, they were motivated to learn digital banking after they felt envy when seeing people around them use it or received electronic money transferred from others.
Although ``social influence'' mostly played a positive role in increasing the adoption of digital banking among older adults, few participants also mentioned that their family members tried to steer them away from digital banking to avoid potential financial loss. This raises an open question of how to help older adults adopt digital banking while minimizing potential financial loss. 

In general, our participants felt that they might be slow in learning or need some help along the way just as the previous research pointed out~\cite{czaja2006factors, olson2011diffusion,Smith2014Older, van2014revisiting,yu2020maps,mohadisdudis2014study,khawaji2017overcoming}, but they believed they could still do it. They also felt proud of being able to use digital banking.}

\subsection{Challenges in Using Digital Banking}
We discuss design considerations to address the challenges older adults encounter when using digital banking.

\subsubsection{App-Related Challenges} 
We uncovered four types of app-related challenges: \textit{ambiguous affordance}, \textit{low information scent}, \textit{insufficient error recovery support}, and \textit{lack of feedback or confirmation}. 
The first two challenges were also found in a recent study about older adults using digital maps~\cite{yu2020maps}. The common challenges shared across drastically different apps (i.e. digital banking in our case vs. mobile maps~\cite{yu2020maps}) suggest that mobile app designs suffer from common design flaws for older adults.

Furthermore, modern app designs also tend to use icons to indicate functions~\cite{SystemIc75:online}. Although such designs can reduce text, our study suggests that icon-based designs may lead to \textit{ambiguous affordance} issues. Instead of just focusing on solving a particular type of app issue, we, as a community, should conduct more research to carefully scrutinize whether general mobile app design practices and guidelines are suitable for older adults, and more importantly how they might need to be adapted based on older adults' experiences. Toward this goal, we discuss some potential design solutions. 

\textbf{Minimize the number of visual elements per screen}.  When reviewing digital banking apps, we noticed that they tended to show many options on one screen. Although app designers often use different font sizes and colors to provide grouping and hierarchical information to distinguish different options (e.g., ~\cite{ColorVis12:online, Navigati64:online}), which is a good practice in general, our study and other ones together provide evidence that some UI elements with less visual salience would likely go unnoticed if there are too many visual elements. As a result, reducing the number of visual elements per screen would likely increase the chance of each element being noticed by older adults. What functions should be shown on the screen? Previous research shows that older adults might prefer a multi-layered interface design, which provides reduced functions for initial learning and then progressively increases the interface complexity, over a full-function interface~\cite{leung2010multi}. However, one challenge with multi-layered interface design is how to present additional functions to older adults. One common approach is to hide them in deeper layers, which will result in a ``deep and narrow'' design. This raises the ``discoverability'' issue. Indeed, our participants mentioned that they did not realize the existence of certain functions that required many steps to reach. It remains an open question of how to balance the ``visual salience'' and ``discoverability'' of a function in mobile apps.

\textbf{Understand older adults' interpretations of common icons}. One cause of the ambiguous affordance issue was the misinterpretation of graphical icons. For example, although a triangle icon might be interpreted as clickable by young adults, our participants did not always perceive it as clickable. \rv{Indeed, Berget G. and Sandnes F.E. found that icons were not universally known by all, and age had a positive correlation on the recognition of the aged icons vs timeless icons~\cite{Berget15On}. Older adults were found to have more problems using existing mobile device icons than younger adults~\cite{Leung2011AgerelatedDI}. A recent study also showed that older adults might misinterpret icons and interactive elements in online visualizations~\cite{Fan2022Visualization}. Furthermore, Rock et al. summarized that four icon characteristics---semantically close meaning (i.e. natural, a close link between depicted objects and associated function), familiar, labeled and concrete (i.e. those depicting real-world objects)---improved its usability for older adults, and they suggested allowing users to choose an icon from a set of potentially suitable icons~\cite{Leung2011AgerelatedDI}. Similar to icons, we also found misinterpretations of mobile app UI elements. 
Understanding how older adults perceive icons or common mobile app UI elements would provide insights for app designers to design ones that match older adults' expectations. Thus, future work should systematically explore older adults’ interpretation of banking icons or UIs. \camera{In addition to universal icons, icons with a specific cultural element can also be designed for older adults living in that culture.}}

\textbf{Explore interactive intelligent agents to assist older adults}. One way to combat the ``low information scent'' and ``insufficient error recovery support'' is to allow older adults to ask questions when they are confused. As voice-based intelligent assistants (e.g., iOS Siri~\cite{Siri}, Amazon Echo~\cite{Echo}, Google Home~\cite{GoogleHome}) continue being improved, researchers began to investigate their potential in assisting older adults~\cite{kim2021exploring,shalini2019development}. Indeed, our participants also expressed the need to ask others when learning digital banking. Thus, it is worth investigating whether and how to design voice-based intelligent assistants to help older adults overcome challenges when using mobile apps (e.g., banking) when they have no idea what to do with an interface (low information scent) or need to recover from an error. Toward this goal, older adults' think-aloud verbalization and voice features can be used to build AI models to detect when they encounter problems~\cite{Fan2021OlderAdults,Fan2020Automatic,Fan2019Concurrent}. Furthermore, bank employees mentioned that some pioneer banks started to add voice interaction into mobile banking apps. However, for older adults with poor hearing, the devices could not automatically increase their voice to an appropriate level and react like a real bank employee. Future work should explore how to design voice assistants that are not only able to detect the problems older adults encounter but also interact with them like a real bank employee.

\camera{
\textbf{Combination of human support and advanced technologies.}
In addition to the aforementioned automatic voice assistants with intelligence and voice-based interfaces, we can also consider combining automatic features with human support to help older adults. For example, when older adults first learn to use digital banking, it might be more helpful for them to interact with a real remote bank teller, who can empathize better than a pure automatic agent. Recent technologies, such as virtual and augmented reality, might also allow older adults to interact with a live remote bank teller through embodied avatars and shared visual cues. Moreover, future research could explore ways to design features that enable help-givers to draw visual guidance on older adults’ phones to support them in conducting banking transactions and even support them to perform trial-and-error, which is effective for learning new technologies but is often challenging for older adults to do~\cite{Fan2018Guidelines,Fan2022Visualization}. Another approach is to design a remote collaborative tool that allows help-givers to demonstrate how to complete a banking task with interactive guidance for older adults.}


\subsubsection{User-Related Challenges} 
\label{discussion:user_challenges}
We propose the following design considerations to address three types of user-related challenges: \textit{anxiety}, \textit{low perceived self-efficacy},  and \textit{memory concerns}. 

\textbf{Increase older adults' understanding of digital banking \rw{with social help}}. The main causes for \textit{anxiety} were related to the abstraction, trustworthiness, and security and privacy concerns of digital banking platforms. \rw{Previous research found that older adults are concerned about security(e.g., ~\cite{Mitzner2010Older, Lian2014Online}) and privacy(e.g., ~\cite{McNeill2017Function}) issues when they access digital technology such as smartphones and health monitor systems and they also experience challenges with managing online security behaviors (e.g., ~\cite{Jiang2016Generational,Anderson2010PracticingSC,Frik2019Privacy,Nicholson2019If}) and privacy settings (e.g., ~\cite{Frik2019Privacy,Hor2017Navigating,Nicholson2019If}) on their own. More recently, the social support approach provided by older adults’ social networks especially close-ties either family or peer friends has been found effective to help older adults manage the security and privacy of smartphones~\cite{Mendel2019Social,Wan2019AppMoD,Kropczynski2021Towards}. Compared to general smartphone usage, digital banking is more sensitive to older adults with possible economic risks and privacy leaks, which may lead older adults to feel more anxious while using. Thus, more research is warranted to investigate ways to better inform older adults how digital banking works, help them debunk false impressions and build confidence with social support while considering how to avoid invasion of privacy during the support.}

In the meantime, banking entities should also pay more attention to security issues and further improve the mechanism of vulnerability patching (e.g., provide various channels to collect flaws reported by users)~\cite{chen2018mobile} to fundamentally alleviate older adults' anxiety about digital banking platforms.

\textbf{Integrate the feeling of companionship into the design of digital banking}. On the other hand, concerns about digital banking's security and privacy were not unique to older adults. \rv{Younger adults were also shown to worry about the perceived risk and privacy of digital banking services~\cite{Laforet2005, Hua2008}. One potential reason was that online services lack physically present security personnel as in physical banks~\cite{Krisnanto2018Digital}.} This suggests the importance of \textbf{the companionship of a trustworthy person for creating a safety feeling} of digital banking platforms. This was echoed by both the fact that older adults would rather wait a long time to receive assistance from bank employees~\cite{Xiaofu21} and in bank employees' observations that older adults felt safe when talking to them and receiving printed receipts of the transactions. The sympathy that helps givers express when interacting with older adults is what current digital services lack, as S6 articulated: \textit{``Machines are not as smart as bank employees, but more importantly, not as sympathetic as them.''} However, bank employee participants felt that it was challenging to keep being sympathetic and patient when they had too many customers waiting to be served. One potential future direction perhaps is to integrate companionship and sympathy into the design of voice-based assistants to not only offer help but also express sympathy and companionship.

\textbf{Increase older adults' perceived self-efficacy}. Our study found that the main causes for ``low perceived self-efficacy'' were \textit{being afraid of making mistakes} and \textit{lack of confidence in their literacy level and their declining physical conditions}. One common underlying factor was the \textbf{potential financial loss associated with mistakes or misoperation}. This suggested that older adults might be more sensitive to \textbf{``loss aversion''}~\cite{kahneman2011thinking} when dealing with digital banking platforms. This was evident in the approaches participants took to minimize their potential financial loss: set an upper limit for the amount of money put in digital banking platforms; bind a dedicated non-primary bank account with the digital banking platforms; practice with a small amount of money. Future work should investigate ways to help older adults cope with loss aversion by boosting their confidence when using digital banking platforms.

One potential approach is to \textbf{deliver multi-model confirmation} instead of just visual confirmation. Current mobile app confirmation design heavily depends on visual feedback. However, older adults often have declining eyesight and may not be sensitive to all sorts of visual feedback (e.g., popup boxes). Instead, the app could deliver the confirmation in multiple channels, for example, by popping up a box and reading the confirmation out \cite{lee2009effect}. 

Another approach is to design better \textbf{error recovery mechanisms}. Current error-recovery mechanisms often require users to accurately press a series of buttons (e.g., a back button placed on the top left corner) to return to a previous step. However, it is not uncommon for older adults to encounter touch-related motor issues~\cite{culen2013touch, peters2016losing, petrovvcivc2018design}, such as tapping and swiping. Thus, touch-based error recovery mechanisms might even lead to more errors along the way. Furthermore, it is also challenging for older adults to figure out which step they should backtrack to. This raises an open question of how best to help older adults recover from an error. One possible direction is to leverage artificial intelligence to infer the step where older adults start to deviate from a correct completion path and later guide them to backtrack to the step. 

Lastly, our findings revealed common ``memory concerns'', such as forgetting passwords or to do a required action. There is a body of literature investigating ways to alleviate these memory issues, such as designing more memorable authentication mechanisms (e.g.,~\cite{renaud2007now}) and reminder systems (e.g.~\cite{chan2019prospecfit,li2019fmt}).  
Moreover, participants were also afraid of forgetting the steps of completing a bank transaction and wrote the steps down in a notebook. However, \textit{writing the steps down} is not a scalable approach to learning. One potential solution is to leverage AI technology to \textbf{learn the steps for completing a task} and alleviate older adults from needing to remember the steps. For example, Li et al. showed an approach for the mobile app to learn from the user how to complete a task and then automate the process for the user for the same and even other similar tasks~\cite{li2017sugilite}. Future work could investigate similar approaches to help older adults complete tasks without needing them to remember the exact steps.

\section{Limitations and Future Work}

Our study presents a qualitative understanding of why and how older adults in China use both physical and digital banking platforms, how they learn to use digital banking platforms and the challenges that they encounter. Although our study included older adults from rural areas (e.g., third-tier and fourth-tier cities), the majority of the participants lived in middle- and big-sized cities. As the general technology development and the availability of digital banking services might differ in different-sized cities, our findings might not reflect the banking practices and challenges of older adults living in regions with different levels of economical development. 

China has been experiencing one of the fastest growth in electronic transactions. Thus, older adults in China might have felt stronger peer pressure to adopt digital banking compared to countries where digital banking is yet to gain popularity. Moreover, the culture, household income, the occupational and educational background of older adults may also affect their mindsets about money and their financial management strategies. Future work should investigate older adults' banking practices in different countries to better understand cultural impacts and derive common and unique challenges. 

Furthermore, our study did not compare the user experience of different banking apps. A controlled experiment with older adults using different banking apps would be able to understand the pros and cons of different app designs. 

Last but not least, our study shows that older adults are not helpless but instead are quite willing to learn new digital banking technologies and apply their accumulated experiences and knowledge to come up with creative solutions before seeking help from someone whom they trust. Inspired by this finding and the advocate for positive aging~\cite{gergen2001positive,Knowles2021TheHarm}, we argue that we should design interactive learning approaches \textit{with} older adults so that they could play a more active and leading role in exploring new technological solutions that still allow them to gain support from their close ties (e.g., children, grandchildren); This could also be a great opportunity to foster a stronger intergenerational bind, which might potentially reduce ageism~\cite{marques2020determinants}.



\bibliographystyle{ACM-Reference-Format}
\bibliography{main}










\end{document}